\documentclass[useAMS,usenatbib,usegraphicx, a4paper]{mn2e}

\usepackage[total={17.8cm,24.0cm},centering]{geometry}
\usepackage{times}
\usepackage{deluxetable}
\newcommand{\aj}{{AJ}}                  
           
\newcommand{\apj}{{ApJ}}                        
\newcommand{\apjl}{{ApJ Lett.}}         
\newcommand{\apjs}{{ApJS}}              
            
\newcommand{\aap}{{A\&A.}}

\newcommand{\mnras}{{MNRAS}}

\newcommand{\pasp}{{PASP}}              
\newcommand{\pasj}{{PASJ}}

\newcommand{\nat}{{Nature}}

\title{ Statistical Properties of Brown Dwarf Companions: Implications for 
 Different Formation Mechanisms}

\author{Bo Ma$^{1}$\thanks{E-mail:boma@astro.ufl.edu} and 
Jian Ge$^{1}$ \\
$^{1}$Department of Astronomy, University of Florida, 211 Bryant Space Science Center, Gainesville, FL, 32611-2055, USA
}

\begin{document}

\maketitle

\label{firstpage}

\begin{abstract}

The mass domain where massive extrasolar planets and brown dwarfs 
overlap is still poorly understood due to the paucity of brown dwarfs orbiting 
close to solar-type stars, the so-called brown dwarf desert. 
In this paper we collect all of available data about close brown dwarfs 
around solar type stars and their host stars from literature and study the 
demographics of the brown dwarf desert. The data clearly show a short 
period and a medium mass gap in the brown dwarf period-mass distribution diagram 
($ 35<m\sin i<55 M_{\rm Jup}$ and $P<100$ days), representing the 
``driest land'' in the brown dwarf desert. Observation biases are highly unlikely 
to cause this gap due to its short period and medium mass, of which 
brown dwarfs can be easily detected by previous RV surveys. 
Brown dwarfs above and below this gap have significantly different 
eccentricity distribution, which not only confirms that this gap is real, but 
also implies that they may have different origins. Our further statistical 
study indicates that brown dwarfs below this gap may primarily form 
in the protoplanetary disk through disk gravitational instability, while 
brown dwarfs above this gap may dominantly form like a stellar binary 
through molecular cloud fragmentation. Our discoveries have offered 
important insights about brown dwarf formation mechanisms and 
their possible relationships with planet and star formation. 
\end{abstract}

\begin{keywords}
stars: brown dwarf -- technique: radial velocity
\end{keywords}

\section{INTRODUCTION} 

Brown dwarfs (BD) are in the mass range of approximately 13$-$80 Jupiter 
masses, having sufficient masses to burn deuterium but not enough to 
burn hydrogen in their inner cores \citep{burrows97, chabrier00, burrows01, spiegel11}. 
The first discovery of a bona-fide BD (Rebolo et al. 1995; Nakajima et al. 1995; 
Oppenheimer et al. 1995; Basri et al. 1996; Rebolo et al. 1996) came in the same 
year as the discovery of the first extra-solar planet around a solar type star, 
51 Peg b \citep{mayor95}. 
One of the major achievements of high-precision radial velocity (RV) surveys over the 
past two decades is the identification of a brown dwarf desert, 
a paucity of brown dwarf companions relative to planets within 3 AU 
around main-sequence FGKM stars (Marcy \& Butler 2000; Grether \& Lineweaver 2006). 
Although the induced reflex RV signal by a close BD companion on a solar type star 
is well within the detection sensitivities of the high precision 
RV surveys ($\sim 3-10$ m/s), only a few dozens are known 
(Reid \& Metchev 2008; Sahlmann et al. 2011a and references therein) compared 
to over five hundred giant planets detected so far by RV technique. 
The California \& Carnegie Planet Search measured an occurrence rate of 
0.7\% $\pm$ 0.2\% from their sample of $\sim \! 1000$ target stars 
\citep{vogt02, patel07}, and the McDonald Observatory Planet Search 
shows a similar rate of 0.8\% $\pm$ 0.6\% from a search sample of 250 
stars \citep{wittenmyer09}. Sahlmann et al. (2011a) obtained an upper 
limit of $0.6\%$ for the frequency of close BD companions based on 
the uniform stellar sample of the CORALIE planet search, which 
contains 1600 solar type stars within 50 pc.

To assess the reality of the brown dwarf desert, \citet{grether06} performed 
a detailed investigation of the companions around nearby Sun-like stars. 
They found that approximately $16\%$ of nearby Sun-like stars have close ($P<5$ yr) 
companions more massive than Jupiter: 11$\%\pm3\%$ 
are stellar companions, $<1\%$ are BDs, and $5\%\pm2\%$ are giant planets.
Although the close BDs are rare around solar type stars, 
\citet{gizis01} suggests that BDs might not be as rare at wide 
separations \citep[see also][]{metchev04} as at close separations. 
\citet{laf07} obtained a $95\%$ confidence 
interval of $1.9^{+8.3}_{-1.5}\%$ for the frequency of 13$-$75$M_{\rm Jup}$ 
companions between 25$-$250AU in the Gemini Deep Planet Survey 
around 85 nearby young stars. \citet{metchev09} inferred the frequency of BDs
 in 28$-$1590 AU orbits around young solar analogs is $3.2^{+3.1}_{-2.7}\%$ 
from an adaptive optics survey for substellar companions around 266 Sun-like stars.

BDs are traditionally believed to form like stars, through gravitational collapse 
and/or fragmentation of molecular clouds \citep{padoan04, hennebelle08}. 
A recently found self-gravitating clump of gas and dust has a mass ($0.015-0.03 \; M_\odot$) 
in the BD regime \citep{andre12}, which supports the idea that BDs could form like a star.
On the other hand, companions with masses up to 10 $M_{\rm Jup}$ \citep{alibert05} 
or even 38 $M_{\rm Jup}$ (Mordasini et al. 2009) may form in 
protoplanetary disks according to the standard core-accretion planet formation theory. 
Because of this, the brown dwarf desert is commonly interpreted as the gap between the largest 
mass objects that can be formed in protoplanetary disks, and the smallest 
mass clumps that can collapse and/or fragment in the vicinity of a protostar. 
The mass function of close stellar companions shows a linear decrease in 
log(M) toward the BD mass range from both stellar mass and planetary mass
directions (Grether \& Lineweaver 2006). In comparison, the mass function of isolated 
substellar objects seems to be roughly flat in log(M) down to masses 
$\sim 20\; M_{\rm Jup}$, both in the field and in clusters (Luhman et al. 2000; Chabrier 2002). 
This indicates that close BD companions may form in a different way from 
those formed in the field and clusters.  

As such, statistical properties of close BD companions 
as well as how these properties are related to their host stars, 
contain a lot of information about the poorly understood BD formation mechanisms and 
their relationships with star and planet formations in close orbital environments.
 These statistics may also be important to investigating how additional 
important processes such as tidal evolution and disk-planet 
interaction affect close BD properties \citep[e.g.,][]{armitage02,matzner05}. 
Given that the close BD occurrence rate is $<1\%$, currently there has  yet been a single 
large, relatively uniform RV survey capable of producing a large homogeneous sample of 
BD companions  for a meaningful statistical study (Marcy et al. 2000; Ge et al. 2008; 
Sahlmann et al. 2011a). However, all of the previous RV planet surveys have
sufficient RV sensitivity and time baseline (at least more than 2 years, e.g., 
Ge et al. 2008; Ge et al. 2009; Eisenstein et al. 2011) to detect a majority of close 
BDs around solar type stars due to the large RV amplitude (on the order 
of $\sim$ 1000 m/s, vs. a few to a few tens m/s RV precisions in previous 
RV planet surveys). It is, therefore, possible to combine close BD 
companion samples together for a statistical study without major biases. 

In this paper we assembled a catalog of all the BD companions discovered 
around solar type star from literature and used it to conduct a statistical study. 
We have found tentative evidence for the existence of two different 
populations of BD companions. 
We present the BD catalog assembled in this study in \S 2 , 
statistical properties of BD companions in \S 3 and discuss their 
implication for different BD formation scenarios in \S 4. We summarize our main results  in \S 5.

\section{Catalog Description} 
We have collected  data from literature about the currently known BD (candidates) 
companions around FGK type stars. Most of them 
have known Keplerian orbits, except for HIP 78530 \citep[discovered by 
direct imaging;][]{laf11}, KOI-205.01 \citep[unpublished yet;][]{santerne12} 
and KOI-554.01  \citep[unpublished yet;][]{santerne12}. 
Properties of the BDs (minimum mass, period and eccentricity) and their host stars  
(mass, effective temperature, surface gravity and metallicity) are summarized 
in Table~\ref{tab:catalog}. For those which are transiting their parent stars or 
have astrometry measurements, true masses are also given besides $M\sin i$.

\begin{figure}
\begin{center}
\includegraphics[width=8cm,trim=1cm 0.7cm 1cm 0cm]{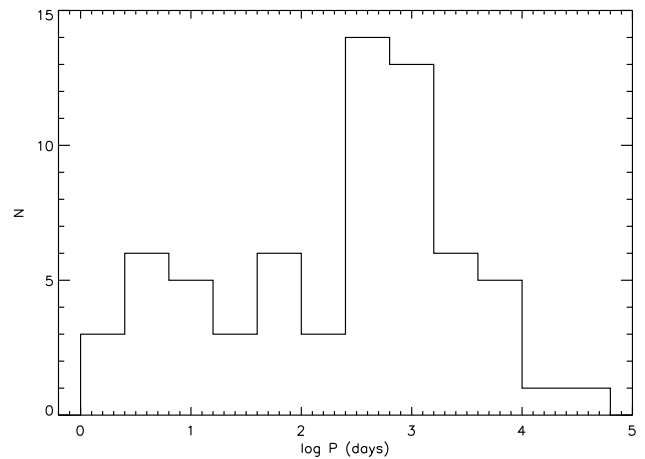}
\caption{\label{fig:period} 
Period distribution of known BD companions around solar type stars.
 }
\end{center}
\end{figure}

\begin{figure}
\begin{center}
\includegraphics[width=8cm,trim=1cm 0.7cm 1cm 0cm]{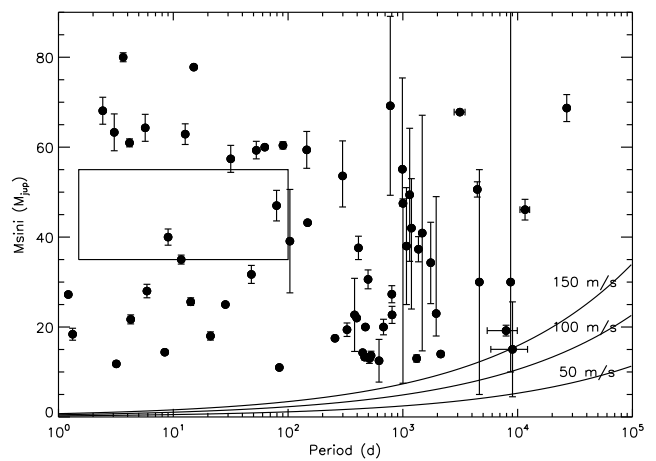}
\caption{\label{fig:gap}
Cumulative mass distribution of brown dwarf candidates. Three lines with 
three RV precisions, 50 m/s, 100 m/s and 150 m/s, are also shown. }
\end{center}
\end{figure}

\section{Observed properties of Brown Dwarfs}
\subsection{Orbital Period Distribution}
The distribution of orbital periods of BDs has two 
main features (Fig.~\ref{fig:period}): a relatively flat distribution inside $P\sim100$ 
days and a sharp jump beyond $P\sim100$ days. The drop beyond $P\sim1000$ days 
is likely due to the observational incompleteness since 1) it is more difficult to detect a BD companion over
a long  period than a short period with RVs and 2) some RV surveys 
(such as the SDSS-III MARVELS, Ge et al.  2008; Ge et al. 2009;
Eisenstein et al. 2011) do not cover beyond this period.  It is evident that the 
number of BDs increases with the orbital period, even though RV and transit observations 
are biased toward discovering objects in short-periods. 
The position of the maximum of the distribution is unknown due to 
the different duration limit of most of the old surveys (several thousand days).
This increasing distribution is consistent with the results from high 
contrast and high angular resolution imaging surveys, 
which find evidence for a higher fraction of BD companions at wide orbits than at close orbits 
\citep{laf07, metchev09, chauvin10, janson12}.

For comparison, both extrasolar giant planets \citep{cumming99, udry03, marcy05, udry07} 
and binaries \citep{duquennoy91} show an increasing number distribution with period. 
However, there is no 3-day pile up with the BD distribution as shown in 
giant planets (Udry et al. 2003). The reason may be that BDs initially form in 
the protoplanetary disks further away from the host stars where protoplanetary disks have more 
materials  to efficiently form BDs than at the close-in regions, and the migration 
mechanism may not efficiently move such a massive body to a short period orbit 
(Trilling et al. 1998; Nelson et al. 2000; Trilling, Lunine \& Benz 2002). 
On the other hand,  BDs forming at the same 
time as the primary stars may migrate inward quickly in the initial gas rich disks and 
thus be destroyed via mergers with the stars \citep{armitage02}.

\begin{figure}
\includegraphics[width=8cm,trim=1cm 0.7cm 1cm 0cm]{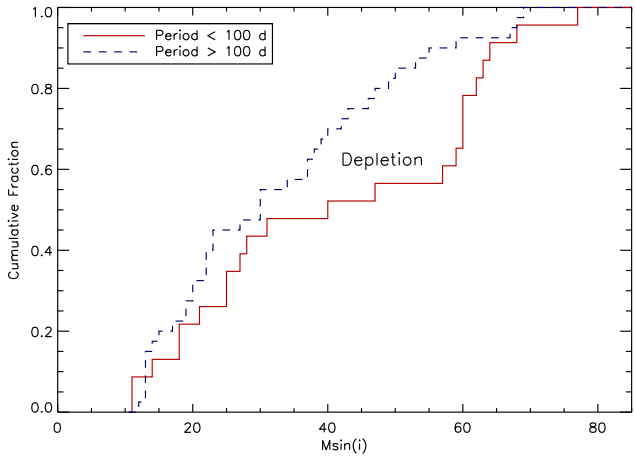}
\caption{\label{fig:mass_cumu}
Cumulative mass distribution of brown dwarf companions. Brown dwarfs with periods 
greater and less than 100 days are shown as dashed and solid lines respectively. }
\end{figure}

\subsection{The Period-Mass Diagram}

The period-mass distribution of BDs shows a statistically significant gap at the 
short period and medium mass region as illustrated in  Fig.~\ref{fig:gap}. 
In this plot, a rectangular area with $P<100$ d and 
$35 M_{Jup}<M<55M_{\rm Jup}$ is highlighted to show the gap 
within which BDs are nearly depleted while there are numerous BDs 
around this region. It appears that this gap is 
real since it is unlikely caused by detection sensitivity (RV precision) or 
observation biases (due to survey incompleteness). 
Previous RV observations have detected many BDs with masses less than the lower limit 
($\sim$35 M$_{Jup}$) of this region and also BDs with periods significantly 
longer than the period limit of this region ($\sim$100 days). RV sensitivities 
with three moderate RV precisions, 50 m/s, 100 m/s and 150 m/s, shown 
in Fig.~\ref{fig:gap} clearly illustrate that any of BDs in the gap should be detected 
easily with these moderate RV precisions. To further verify if this feature 
is real, we divided the BD sample into two groups according to their periods 
($P<100$ days and $P>100$ days) and plotted their mass cumulative 
histograms in Fig.~\ref{fig:mass_cumu} for comparison. It is apparent that 
a depletion of BDs with masses between 35 and 
$55M_{\rm Jup}$ appears in the cumulative histogram for the short 
period group while no depletion of BDs appears in the cumulative 
histogram for the long period group. We carried out a simple 
Monte Carlo experiment to test the emptiness of this gap 
on the period-mass diagram. There are a total of 25 BDs with period shorter than 
100 days in our BD sample. We assumed simply that their masses are 
uniformly distributed between 13 and 80$M_{Jup}$. Then we drew their masses 
randomly from this uniform distribution and counted how many of them will fall 
in the mass range of 35 to $55M_{\rm Jup}$. We found the probability 
that less than 3 BDs will fall in this gap is $0.9\%$, which corresponds 
to a $2.6\sigma$ significance. A larger BD sample in the future will 
be better to assess the significance of this gap.

The appearance  of this depleted region in the BD period-mass diagram has naturally 
divided BDs into two mass groups:  one with masses greater than $42.5M_{\rm Jup}$  and the
other less than $42.5M_{\rm Jup}$. Their properties and origins may be different. 
We further explore properties of these two groups and study possible origins. 

\subsection{Orbital Eccentricity Distribution \label{sec:ecc}}

\begin{figure}
\includegraphics[width=8cm,trim=1cm 0.7cm 1cm 0cm]{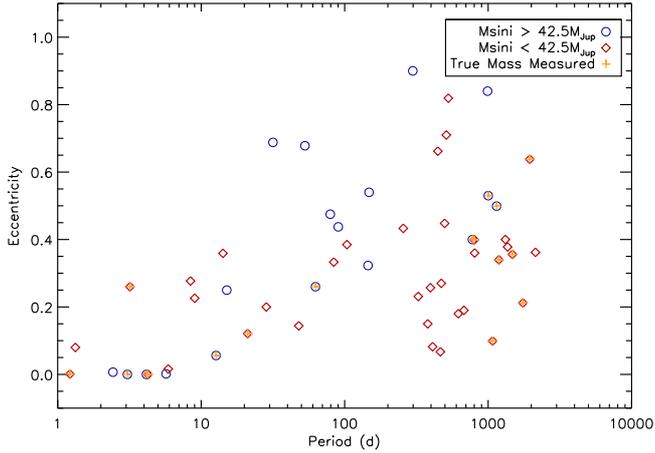}
\caption{\label{fig:p-e}
Period-eccentricity distribution of brown dwarf candidates. Brown dwarf candidates 
with masses above and below 42.5$M_{\rm Jup}$ are shown as circles and diamonds. 
Brown dwarf candidates with true masses measured using transiting observations 
or astrometry measurements are shown as crosses.}
\end{figure}

The orbital eccentricities show great difference for the two BD groups 
with masses greater and lower than 42.5$M_{\rm Jup}$,
respectively. Fig.~\ref{fig:p-e} shows the period-eccentricity distribution of all 
 known BDs.  
The period-eccentricity distribution of BDs with masses greater than $42.5M_{Jup}$ 
is consistent with a circularization limit of $\sim12$ days, which is similar to 
that found in nearby stellar binaries \citep{rag10}.
It is clear that there are a significant number of 
BDs with $300\;{\rm d} <P< 3000 \; \rm d$ and 
$e<0.4$ for BDs with masses lower than 42.5$M_{\rm Jup}$, but no BDs 
with masses greater than 42.5$M_{\rm Jup}$. We have done a two-dimensional 
Kolmogorov-Smirnov (K-S) test for the period-eccentricity distribution of BDs with 
masses greater and lower than 42.5$M_{\rm Jup}$. The probability that these 
two BD samples are drawn from the same distribution 
in the period-eccentricity plane is $1.7\%$. 

Next we are going to compare the period-eccentricity distribution of the BDs to that 
of stellar binaries. \citet{halbwachs03} have studied the statistical properties of a 
sample of 89 FGK type main-sequence binaries with periods up to 10 year. Here 
we choose to use their 89 binary sample for our comparison.
To compare the period-eccentricity distribution between BDs and stellar binaries
we made use of a two-dimensional K-S test. The probability 
for the period-eccentricity distribution to be the same is $18\%$ between 
BDs with masses above 42.5$M_{\rm Jup}$ and the stellar binary sample, 
and $0.1\%$ between BDs with masses below 42.5$M_{\rm Jup}$ 
and the stellar binary sample. These results suggest that BDs with masses 
greater than 42.5$M_{\rm Jup}$ have a very similar period-eccentricity 
distribution to that of stellar binaries, and their formation mechanisms 
may be similar.

\begin{figure}
\includegraphics[width=8cm,trim=1cm 0.7cm 1cm 0cm]{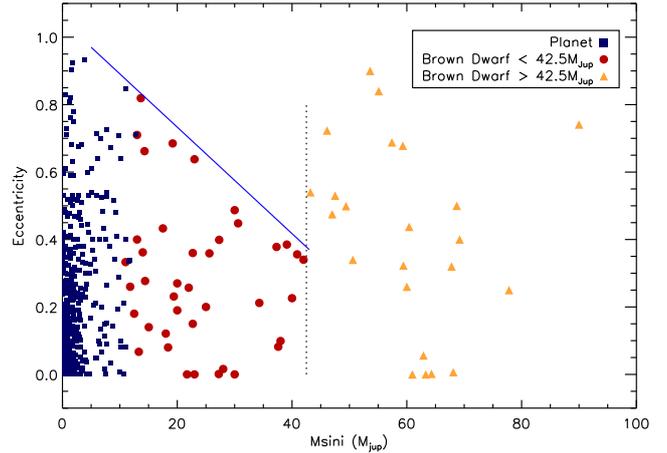}
\caption{\label{fig:m-e}
Mass-eccentricity distribution of exoplanets and brown dwarf candidates. 
The solid line shows a trend that the more massive the giant planet/BD is, 
the smaller maximum eccentricity it tends to have, which breaks at 
$\sim42.5M_{\rm Jup}$. The vertical dotted line shows 
$M\sin i = 42.5M_{\rm Jup}$. Planets are shown as squares. 
Brown dwarfs with masses above and below $42.5M_{\rm Jup}$ 
are shown as triangles and circles respectively. }
\end{figure}

The difference of the orbital eccentricities for BDs with different masses is 
further illustrated in the mass-eccentricity plot shown in  Fig.~\ref{fig:m-e}. By including
all  currently known planets (from exoplanet.org) and brown dwarfs (this paper) in this plot, 
a clear trend is shown:  all the known giant planets and BDs with masses below 
$\sim 42.5 M_{\rm Jup}$ have the eccentricity distribution following a trend, i.e., 
the more massive the giant planet/BD is, the lower maximum eccentricity 
it tends to have while BDs above this mass threshold do not show 
such a trend, instead showing more diversity in their eccentricities.

\subsection{Metallicity of the BD Host Stars}

The BD host stars in this study have a mean metallicity 
of [Fe/H] $= -0.04$ with a standard deviation of 0.28. 
For comparison, \citet{rag10} has carefully studied a sample 
of 454 nearby solar type stars, which have a mean metallicity of $-0.14$ with a 
standard deviation of 0.25. Six stars in that sample have BD
companions, which have a mean metallicity of [Fe/H]$=-0.05$.
The mean metallicity of our BD sample is slightly higher 
than the mean metallically of volume limited 
nearby FGK dwarf stars. For example, \citet{favata97} have analyzed a volume limited sample 
of 91 G and K dwarfs, yielding a mean metallicity of [Fe/H]$=-0.08$ with a 
standard deviation of 0.26. \citet{nordstrom04} have derived metallicity 
for 16682 nearby F and G dwarf stars 
with a mean of $-0.14$ and a dispersion of 0.19 dex.
Sousa et al. (2011) found that a mean metallicity for the CORALIE survey 
sample of 1248 stars and the HARPS survey sample of 582 stars is [Fe/H]=$-0.11$ 
and $-0.10$, respectively. However, since some exoplanet surveys choose 
samples biased towards metal rich stars (e.g., Valenti \& Fischer 2005), the slightly higher 
mean metallicity of BD host stars is possibly caused by the sample bias.

After comparing the BD host star metallicities with the volume limited sample from
\citet{sousa11} and the planet search sample from Valenti \& Fischer (2005), 
we find that the BD host star metallicity distribution is consistent with the combination of these 
two samples. This means we cannot interpret the BD host star sample as metal rich. 
We compared  metallicities of BD host stars with that of giant planet 
($1M_{\rm Jup}<m\sin i<5M_{Jup}$) host stars. The data of giant planet 
host stars is taken from the Exoplanet Orbit Database \citep{wright11}. 
A K-S test shows that the probability of these two samples selected from the same 
distribution is $2\times10^{-4}$. This indicates that the two samples are 
significantly different from each other.

We investigated the correlation between BD host star metallicities and BD masses. 
The Spearman's rank correlation coefficient of the BD mass 
and their host star metallicity is 0.07 with a $61\%$ significance, suggesting there is no 
significant correlation between the BD mass and their host star metallicity.

\begin{figure}
\includegraphics[width=8cm,trim=1cm 0.7cm 1cm 0cm]{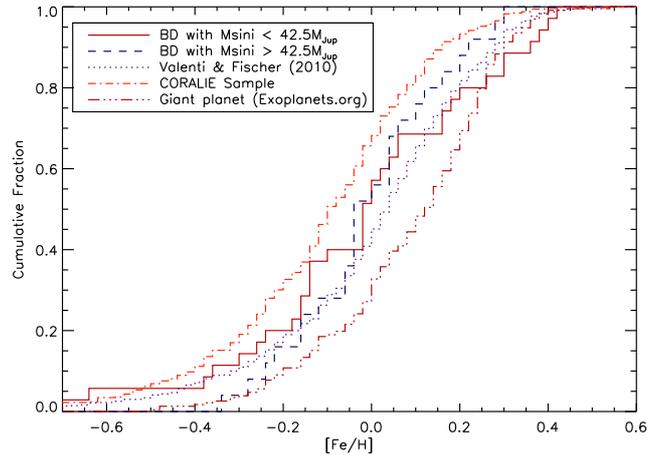}
\caption{\label{fig:meta_cumu}
Cumulative metallicity distribution of host stars with BD companion masses above (dashed line)
and below (solid line) $42.5M_{\rm Jup}$. Also shown for comparison are cumulative metallicity distribution 
of the planet search sample from \citet{valenti05}, the CORALIE planet search sample \citep{sousa11} 
and giant planet sample from the Exoplanet Orbit Database \citep{wright11}.}
\end{figure}

We also divided the BD companions into two subsample according to 
their masses. The cumulative metallicity distribution for host stars with BD companion 
masses above and below $42.5M_{\rm Jup}$ is shown in 
Fig.~\ref{fig:meta_cumu}. The main difference between these 
two distribution is at the lower metallicity end. 
Currently no BDs with masses greater than $42.5M_{\rm Jup}$ have been found around 
stars with [Fe/H]$<-0.5$, while several BDs with masses lower than $42.5M_{\rm Jup}$ 
have been found in this metallicity regime. A K-S test has been conducted to show that the probability of 
these two sample metallicities are drawn from the same distribution is $70\%$, 
which suggests there is no significant difference between these two samples 
regarding their metallicity distribution.

\section{Discussion}

\subsection{Two Different Brown Dwarf Populations}

Our study suggests  that BDs with masses lower than $\sim 43M_{\rm Jup}$ have 
an eccentricity distribution consistent with that of giant planets in the 
mass-eccentricity diagram while BDs with masses above $\sim 43M_{\rm Jup}$ 
have the star-like eccentricity distribution (Figure~\ref{fig:m-e}). Our mass 
limit is consistent with the minimum of the mass functions of planet and 
stellar companions in the BD mass region, $43^{+14}_{-23}M_{\rm Jup}$, 
derived by \citet{grether06} using their stellar companion and giant 
planet sample within 50 pc around the sun. This mass function minimum 
is also consistent with that derived by \citet{sahlmann11b}, who 
found a void in the mass range between 25 and 45 $M_{\rm Jup}$ 
using the data from the CORALIE radial-velocity survey. They further suggested 
that there may be a possible dividing line between massive planets and 
sub-stellar companions. \citet{schneider11} has chosen arbitrarily and probably 
provisionally 25$M_{\rm Jup}$ as the upper limit of massive planets based on previous 
studies \citep[e.g.,][]{sahlmann11b, baraffe10}.

Our BD sample is, therefore, naturally divided into two different groups 
with the mass limit of $\sim 42.5M_{\rm Jup}$. The eccentricity distribution 
of low mass BDs appears to be consistent with the prediction from the 
``planet-planet scattering" model \citep{rasio96, ford08, chatterjee08}, 
while the eccentricity distribution of massive BDs appears to be similar 
to that of stellar binaries \citep{halbwachs03} as shown in \S~\ref{sec:ecc}. 
The eccentricity distributions of 
these two groups support that BDs may form differently: BDs below this mass limit form 
in protoplanetary disks around host stars, and above this mass limit 
form like stellar binary systems. This is supported by our analysis results. 
The existence of a large population of long period low eccentricity 
BDs ($P>300$ days and $e<0.4$) serves as evidence to support the 
BD formation scenario in the protoplanetary disks for those companions 
with masses below 42.5$M_{\rm Jup}$. While the lack of long period 
and low eccentric BD companions with masses above 42.5$M_{\rm Jup}$ 
appears to support the BD formation scenario like stellar binary formation. 
Nevertheless, a small number of BDs in each of these two mass regions 
may form in an opposite formation mechanism, but our sample is not sufficient to 
distinguish these minor groups. 


\subsection{BD formation mechanisms}

Our study of metallicity distribution of BDs appears to support the disk instability 
mechanism \citep{boss97} for those BDs formed in the planetary disks, while 
it is inconsistent with the prediction of core-accretion formation mechanism 
\citep{pollack96, ida04, alibert05}. Currently, there are two hotly contested 
theories about giant planet formations:  core-accretion and disk gravitational
instability. In the core-accretion scenario,  giant planets form more efficiently 
around metal rich planetary disks than metal poor planetary disks because 
the higher the grain content of the disk is, the easier the metal core is to build for 
giant planets \citep{pollack96, ida04, alibert05}. In contrast, the disk gravitational 
instability process allows similar formation efficiency for both metal rich 
and metal poor giant planets. Previous results show strong correlation between 
metallicity and giant planet occurrence rate around solar type stars 
\citep{santos01, valenti05, johnson10} strongly support the core-accretion 
formation scenario for giant planet formation \citep{pollack96, ida04, alibert05}. 
However, our result, showing no correlation between the BD occurrence 
rate and metallicity, appears to support the disk instability mechanism for 
formation of  BDs in planetary disks. 

For stellar companions, there appears to be a 
weak anti-correlation between metallicity and stellar companion occurrence 
rate \citep{rag10}, i.e., lower-metallicity clouds might be more likely to 
fragment to form binary stars. If this applies to massive BDs, then we 
expect this anti-correlation. However, our current sample is too 
small to tell, but is at least not inconsistent with this statistics.

\subsection{Upper Mass limit of Planets}
Our study appears to help to address the upper mass limit of "planets".
Here we define a "planet" as an object formed in a protoplanetary disk from 
the companion formation  point of view. It has been long known that the 
low mass end of brown dwarfs overlaps the high mass end of massive 
planets formed in protoplanetary disks. From theoretical simulations, 
planets with masses up to 10 $M_{\rm Jup}$ \citep{alibert05} or even 
38 $M_{\rm Jup}$ (Mordasini et al. 2009) may form in protoplanetary disks.
However, it is difficult to determine the upper mass limit of planet from observations
since the high mass planets detected may actually be brown dwarfs formed
as low-mass binaries. One way to tackle this problem is to search 
for debris disk around massive planet/BD host star, which support the 
idea that these objects are formed like a planet. \citet{moro10}
have found evidence of debris disk around two massive planet host star (HD 38529 and
HD 202206), which supports formation of planets up to 17 $M_{\rm Jup}$ in
a protoplanetary disk. The other way to resolve this problem is through the study of
mass spectrum of massive planets/BD.
Motivated by the observed mass distribution, this planet upper mass limit
has been set at around $43^{+14}_{-23}M_{\rm Jup}$ \citep{grether06} or
25 $M_{\rm Jup}$ \citep{sahlmann11b, schneider11}.
Analysis conducted in this paper was also able to provide a clue about this upper mass limit.
The cumulative BD mass distribution (Fig.~\ref{fig:mass_cumu}) suggests this limit is
in the range $30-60$ $M_{\rm Jup}$ and mass-eccentricity relation (Fig.~\ref{fig:m-e})
suggests that this limit is around 43 $M_{\rm Jup}$. It appears that the upper limit 
is around  43 $M_{\rm Jup}$, which is consistent with
the result found by \citet{grether06}, but is slightly larger than that from \citep{sahlmann11b}.

\section{Summary}

We have searched literature and presented a catalog of BD companions around 
solar type stars found by radial velocity, transiting and astrometry observations. 
We have studied distribution of different parameters of BD companions around 
solar type star, and found that: \\ 
\\
(1) BD companions have an increasing distribution with period, similar to giant planets 
and low-mass binaries;\\
(2) BD companions are almost depleted at $P<100$ d and $30 M_{Jup}<M<55M_{\rm Jup}$ 
in the period-mass diagram;\\
(3) BD companions with masses below $42.5M_{\rm Jup}$ have eccentricity 
distribution consistent with that of massive planets; \\
(4) BD companions with masses above $42.5M_{\rm Jup}$ have eccentricity 
distribution consistent with that of binaries, which shows the expected 
circularization for periods below 12 days, caused by tidal forces over 
the age of the Galaxy, followed by a roughly flat distribution; \\
(5) Host stars of BD companions are not metal rich, and have significantly 
different metallicity distribution when comparing with host stars of giant planets, 
suggesting a formation scenario at least partly different from the core-accretion scenario.

The distribution of BD and their host star properties presented in this paper 
may lend support to such a picture:
(1) BD companions with masses below $42.5M_{\rm Jup}$ form in a protoplanetary disk 
through the the instability-fragmentation scenario, and their eccentricity is excited 
through scattering with other objects formed in this disk or interactions with disk/third body; 
(2) BD companions with masses above $42.5M_{\rm Jup}$ form like stars, through 
molecular cloud fragmentation, similar to the formation of a stellar binary system.

\section*{Acknowledgments}

This research has made use of the Exoplanet Orbit Database
and the Exoplanet Data Explorer at exoplanets.org. We thank Neil Thomas for 
proof reading our paper.
We acknowledg the support from DoD Coopeative Agreement W911NF-09-2-0017,
Dharma Endowment Foundation and the University of Florida.

\clearpage

\begin{deluxetable}{lccccccccl}
\rotate 
\tablewidth{0pt}
\tabletypesize{\scriptsize}
\tablecaption{Close brown dwarfs (candidate) companions to solar-type stars \label{tab:catalog} }
\tablehead{
\colhead{Object}  &  \colhead{$M_c$} &  \colhead{$M_c \sin i$ } &  \colhead{Period} &  \colhead{Eccentricity} &  \colhead{$M_\star$} &  \colhead{$T_{\rm eff}$} &  \colhead{$\log(g)$ } &  \colhead{ [Fe/H]} & \colhead{References} \\
\colhead{}  &  \colhead{($M_{Jup}$)} &  \colhead{($M_{Jup}$) } &  \colhead{(day)} &  \colhead{} &  \colhead{($M_\odot$)} &  \colhead{(K)} &  \colhead{(cgs)} &  \colhead{} &  \colhead{}
}
\tablecolumns{10}
\startdata
HD 30501 & \nodata & $90\pm12$ & 2073.6$^{+3.0}_{-2.9}$ & 0.741 & $0.81\pm0.02$ & 5223$\pm50$ &4.56$\pm0.10$ & $-0.06\pm0.06$ & 1\\
HD 43848 & $      101.8\pm       15.0$ & $24.5\pm1.6$ &      2354.3$\pm       8.9$ &     0.703 & $     0.89\pm    0.02$ &      5334$\pm       92$ &
      4.56$\pm      0.15$ & $     0.22\pm     0.06$ & 1, 56 \\      
HD 52756 & \nodata & $      59.3^{+ 2.0}_{-1.9}$ &      52.8657$\pm       0.0001$ &  0.6780$\pm0.0003$ & $     0.83\pm    0.01$ &      5216$\pm       65$ &
      4.47$\pm      0.11$ & $     0.13\pm     0.05$ & 1\\
HD 89707 & \nodata   & $ 53.6^{+7.8}_{-6.9}$ &      298.5$\pm      0.1$ &    0.900$^{+0.039}_{-0.035}$ & $     0.96\pm    0.04$ &      6047$\pm       50$ &
      4.52$\pm      0.10$ & $    -0.33\pm     0.06$ & 1\\      
HD 167665 & \nodata & $      50.6\pm       1.7$ &      4451.8$^{+27.6}_{-27.3}$ & $0.340\pm0.005$ & $      1.14\pm    0.03$ &      6224$\pm       50$ &
      4.44$\pm      0.10$ & $   -0.05\pm     0.06$ & 1\\
HD 189310 & \nodata & $  25.6^{+0.9}_{-0.8}$ &      14.18643$\pm0.00002$ &     0.359$\pm0.001$ & $     0.83\pm    0.02$ &      5188$\pm       50$ &
      4.49$\pm      0.10$ & $   -0.01\pm     0.06$ & 1\\      
HD 4747 & \nodata & $      46.1\pm       2.3$ &      11593.2$^{+1118.6}_{-1117.6}$ &     0.723$\pm0.013$ & $     0.81\pm    0.02$ &      5316$\pm       50$ &
      4.48$\pm      0.10$ & $    -0.21\pm     0.05$ & 1\\
HD 211847 & \nodata & $      19.2\pm       1.2$ &      7929.40$^{+1999.1}_{-2500.2}$ & $0.685^{+0.068}_{-0.067}$ & $     0.94\pm    0.04$ &      5715$\pm       50$ &
      4.49$\pm      0.10$ & $   -0.08\pm     0.06$ & 1\\
HD 180314 & \nodata & $      22.0$ &      396.03$\pm      0.62$ &     0.257$\pm0.010$& $      2.6\pm     0.3$ &      4917$\pm       100$ &
      2.98$\pm      0.12$ & $     0.2\pm     0.09$ & 3 \\
HD 13189 & \nodata &  20.0 &      471.6$\pm       6.0$ &     0.27$\pm0.06$& $      7$ &      5000$\pm       100$ &
      2.0$\pm      0.1$ & $      0.0\pm      0.10$ & 4 \\
HD 30339 & \nodata &      77.8 &      15.0778$\pm       0.000$ &     0.25 & $      1.1 $ &      6074$\pm       100$ &
      4.37$\pm      0.10$ & $     0.21\pm      0.10$ & 4\\
HD 65430 & \nodata & $      67.8$ &      3138$\pm       342$ &     0.32 & $     0.78$ &      5183$\pm       100$ &
      4.55$\pm      0.10 $ & $   -0.04\pm      0.10$ & 4\\
HD 140913 & \nodata & $      43.2$ &      147.968$\pm       0.000$ &     0.54 & $     0.98$ &      6048$\pm       100$ &
      4.57$\pm      0.10$ & $    0.07\pm      0.10$ & 4\\
HD 174457 & 107.8$^{+23.8}_{-24.1}$ &  \nodata  &      840.8$\pm     0.05$ &     $0.23\pm0.01$ & $      1.19$ &      5852$\pm       100$ &
      4.08$\pm      0.10$ & $    -0.15\pm      0.10$ & 4 \\
HD 38529c & 17.6$^{+1.5}_{-1.2}$ & $      13.99\pm      0.59$ &      2136.14$\pm      0.29$ &     0.362$\pm0.002$ & $      1.48\pm    0.05$ &      5697 &
      3.94$\pm      0.10$ & $    + 0.27\pm      0.05$ & 5  \\
HD 91669 & \nodata & $      30.6\pm       2.1$ &      497.5$\pm      0.6$ &     $0.448\pm0.002$ & $     0.914^{+0.018}_{-0.087}$ &      5185$\pm       87$ &
      4.48$\pm      0.20$ & $     +0.31\pm     0.08$ & 6\\
11Com & \nodata & $      19.4\pm       1.5$ &      326.03$\pm      0.32$ &     $0.231\pm0.005$ & $      2.7\pm     0.3$ &      4742$\pm       100$ &
      2.31$\pm      0.10$ & $    -0.35\pm     0.09$ & 7 \\
HD 119445 & \nodata & $      37.6\pm       2.6$ &      410.2$\pm      0.6$ &    0.082$\pm0.007$ & $      3.9\pm     0.4$ &      5083$\pm       103$ &
      2.40$\pm      0.17$ & $    0.04\pm      0.18$ & 8  \\
HD 131664 & $  23^{+26.0}_{-5.0}$ & \nodata  &      1951$\pm       41$ &     0.638$\pm0.02$ & $  1.10\pm    0.03$ &      5886$\pm       21$ &
      4.44$\pm      0.10$ & $     +0.32\pm     0.02$ & 9, 56  \\
GJ 595 & \nodata & $      60.0\pm       0.0$ &      62.6277$\pm       0.0001$ &     0.26$\pm0.001$ &    0.28 &      3500$\pm       100$ &
      4.83$\pm      0.10$ & $      0.0\pm      0.10$ & 4 \\
HD 162020 & \nodata & $      14.4\pm  0.04$ &      8.428198$\pm0.000056$ &     $0.277\pm0.002$ & $     0.75$ &      4830$\pm       80$ &
      4.76$\pm      0.25 $ & $    +0.01\pm      0.11$ & 12 \\
HD 168443  & $      34.3\pm       9.0$ & \nodata &      1748.2$\pm       1.0$ &     0.2122$\pm0.0020$ & $      1.01\pm    0.05$ &      5555$\pm       40$ &
      4.10$\pm     0.12$ & $    +0.10\pm     0.03$ & 1, 12, 22 \\
HD 180777 & \nodata & $      25.0\pm       0.0$ &      28.44$\pm     0.01$ &     0.20 &       $1.7\pm0.1$ &      7250 &
      4.34 & $    -0.16$ & 1, 24, 25\\
HD 190228  & $ 49.4\pm       14.8$ &  \nodata  &      1146$\pm       16$ &     0.50$\pm0.04$ &      0.83&      5360$\pm       40$ &
      4.02$\pm      0.10$ & $    -0.24\pm     0.06$ & 1, 26 \\
HD 191760 & \nodata & $      38.17\pm  1.02$ &      505.65$\pm      0.42$ &     $0.63\pm0.01$ &1.28$^{+0.02}_{-0.10}$ &      5821$\pm       82$ &
      4.13$^{+0.05}_{-0.04}$ & $     0.29\pm     0.07$ & 1, 27\\
HD 202206 & \nodata &   17.5 &      256.20 $\pm     0.03$ &     $0.433\pm0.001$ & 1.15 &      5765$\pm       40$ &
      4.75$\pm      0.20$ & $     0.37\pm     0.07$ & 1, 12, 28 \\     
HIP 21832 & $      40.9\pm       26.2$ & \nodata  &      1474.9$\pm       10.2$ &     $0.356\pm0.095$ & $      1.0\pm      0.0$ &      5554$\pm       70$ &
      4.32$\pm      0.10$ & $    -0.63\pm      0.10$ & 17 \\
HD 14651 & \nodata & $      47.0\pm       3.4$ &      79.4179$\pm 0.0021$ &     0.4751$\pm0.0010$ & $     0.96\pm    0.03$ &      5491$\pm       26$ &
      4.45$\pm     0.03$ & $   -0.04\pm     0.06$ & 18\\
HD 30246 & \nodata &       55.1$^{+20.3}_{-8.2}$ &      990.7$\pm       5.6$ &     $0.838\pm0.081$ & $ 1.05\pm    0.04$ &      5833$\pm       44$ &
      4.39$\pm     0.04$ & $     +0.17\pm      0.10$ & 18\\
HD 92320 & \nodata & $      59.4\pm       4.1$ &      145.4$\pm     0.01$ &     0.323 & $     0.92\pm    0.04$ &      5664$\pm       24$ &
      4.48$\pm     0.03$ & $    -0.10\pm     0.06$ & 18\\
HD 22781 & \nodata & $      13.65\pm 0.97$ &     528.07$\pm   0.14$ &     0.8191$\pm0.0023$ & $   0.75\pm    0.03$ &      5027$\pm       50$ &
      4.60$\pm     0.02$ & $    -0.37\pm     0.12$ & 18\\
HD 137510 & 20.0-60.0 & $      27.3\pm       1.9$ &      801.30$\pm      0.45$ &     0.3985$\pm0.0073$ & $      1.36\pm    0.04$ &      6131$\pm       50$ &
      4.02$\pm     0.04$ & $     0.38\pm      0.13$ & 10, 11, 18, \\
HIP 5158 & \nodata & $      15.04\pm    10.55$ &      9017.76$\pm   3180.74$ &     0.14$\pm0.10$ & $     0.780\pm    0.021$ &      4962$\pm   89$ &
      4.37$\pm      0.20$ & $     0.10\pm      0.07$ & 29, 52 \\
HD 41004B & \nodata & $      18.37\pm    0.22$ &  1.328300$\pm   0.000012$ &    0.081$\pm0.012$ & $  0.40\pm    0.04$ &   \nodata &
       \nodata &   \nodata  & 30  \\
HAT-P-13c & \nodata & $      14.28\pm      0.28$ &      446.27$\pm   0.22$ &     0.6616$\pm0.0054$ &    1.22$^{+0.05}_{-0.10}$ &      5640$\pm       90$ &
      4.13$\pm     0.04$ & $     0.430\pm     0.08$ & 31 \\
BD+202457b & \nodata & $      22.7\pm8.1$ &      379.63$\pm       2.01$ &     $0.15\pm0.03$ & $      2.8\pm      1.5$ &      4137$\pm       10$ &
      1.51$\pm     0.05$ & $     -1.00\pm     0.07$ & 32 \\
BD+202457c & \nodata & $      13.2\pm4.7$ &      621.99$\pm       10.20$ &     $0.18\pm0.06$ & $      2.8\pm      1.5$ &      4137 $\pm       10$ &
      1.51$\pm     0.05$ & $     -1.00\pm     0.07$ & 32 \\
HD 137759 & \nodata & $      12.7\pm       1.08$ &      511.098$\pm     0.089 $ &     0.7124$\pm0.0039$ & $      1.80\pm     0.23$ &      4500 $\pm       110 $ &
      2.74$\pm      0.10$ & $    0.03\pm      0.10$ & 11, 33, 35 \\
 NGC 2423-3b & \nodata & $      10.64\pm 0.93$ &      714.3$\pm  5.3$ &     0.21$\pm0.07$ & $      2.4\pm     0.2$ & \nodata &
      \nodata &  \nodata & 36, 37 \\
  NGC 4349$-$127b & \nodata & $      20.0\pm       1.73$ &      678.0$\pm       6.2$ &     0.19 & $      3.9\pm     0.3$ &      4569 $\pm       69$ &
      2.08$\pm      0.35$ & $    -0.13\pm      0.18$ & 36, 37 \\
HD 16760b & \nodata & $      13.13\pm      0.56$ &      466.47$\pm  0.35$ &    0.084$\pm0.003$ & $     0.78\pm    0.05$ &      5629$\pm       44$ &
      4.47$\pm      0.06$ & $   +0.067\pm     0.05$ & 38 \\
HD 10697  & $    38\pm 13$ & \nodata &      1075.0$\pm       1.5$ &    0.099$\pm0.007$ &      1.112$^{+0.026}_{-0.02}$ &      5680 $\pm       44 $ &
      4.12$\pm     0.06$ & $     0.19\pm     0.03$ & 40, 41, 42 \\
HD 114762 & \nodata & $      10.99\pm   0.09$ &      83.9152$\pm0.0028$ &     0.3325& $     0.89\pm    0.09$ &      5950 $\pm       44 $ &
      4.54$\pm     0.06$ & $    -0.65\pm     0.03$ & 11, 34, 43 \\
TYC~2534-698-1 & \nodata & $      39.1\pm       11.5$ &      103.698$\pm      0.111000$ &     0.385 & $     0.998\pm    0.040$ &      5700 $\pm       80 $ &
      4.50$\pm      0.10$ & $    -0.25\pm     0.06$ & 44 \\
TYC~2949-557-1 & \nodata & $      64.3\pm       3.0$ &      5.69449$\pm0.00029$ &   0.0017$^{+0.0019}_{-0.0017}$ & $      1.25\pm    0.09$ &      6135 $\pm       40 $ &
      4.4$\pm      0.1$ & $     0.32\pm     0.01$ & 45 \\
TYC~1240-945-1 & \nodata & $      28.0\pm       1.5$ &      5.8953$\pm  0.0004$ &    0.015$\pm0.011$ & $      1.37\pm     0.11$ &      6186 $\pm       92 $ &
      3.89$\pm     0.07$ & $    -0.15\pm     0.04$ & 16 \\
HIP 67526 & \nodata & $      62.6\pm      0.6$ &     90.2695$\pm0.0188$ &     0.4375$\pm0.0040$ & $      1.11\pm    0.08$ &      6004 $\pm       29 $ &
      4.55$\pm      0.15$ & $    0.04\pm     0.05$ & 63 \\
TYC~2930-872-1 & \nodata & $      68.1\pm       3.0$ &      2.430420$\pm0.000006$ &   0.0066$\pm0.0010$ & $      1.21\pm    0.08$ &      6427 $\pm       33 $ &
      4.52$\pm      0.14$ & $   -0.04\pm     0.05$ & 58 \\
TYC~2087-255-1 & \nodata & $      40.0\pm       1.8$ &      9.0090$\pm    0.0004$ &     0.226$\pm0.011$ & $      1.16\pm    0.08$ &      5903 $\pm       42$ &
      4.07$\pm      0.16$ & $    -0.23\pm     0.07$ & 21 \\
TYC~3130-160 & \nodata & $      57.4\pm       3.0$ &      31.66$\pm     0.023$ &     0.688 & $      1.0\pm     0.1$ &      5104$\pm       50
$ &      4.43$\pm      0.10$ & $    0.01\pm     0.05$ & 62 \\
HIP 78530 & \nodata & $      23\pm       3$ &  $\sim 4.38\times 10^6$  &      0 (fixed) & $      2.5\pm     0.2$ &      10500$\pm  500$ &
     \nodata &  \nodata & 46 \\
HD 5388b  & $      69.2\pm   19.9$ & \nodata &      777.0$\pm       4.0$ &     0.40$\pm0.02$ & $      1.21\pm     0.10$ &      6297$\pm       32$ &
      4.28$\pm     0.06$ & $    -0.27\pm     0.02$ & 48, 53 \\
HR 7672b & $      68.7\pm       3.0$ & \nodata  &      26772$^{+803.5}_{-1059.2}$ &     0.5 & $      1.08\pm    0.04$ &      5883$\pm       59$ &
      4.42$\pm     0.06$ & $    0.05\pm     0.07$ & 49 \\
HD 175679 & \nodata & $      37.3\pm       2.8$ &      1366.8$\pm 5.7$ &     0.378$\pm0.008$ & $      2.7\pm     0.3$ &      4844$\pm       100$ &
      2.59$\pm      0.10$ & $    -0.14\pm      0.10$ & 50 \\
HD 136118 & $42^{+11}_{-18}$ & 12.0$\pm0.47$ & 1188.0$\pm2.0$ & 0.34$\pm0.01$ & 1.24$\pm0.07$ &6097$\pm44$ & 
 4.16$\pm0.09$  & -0.01$\pm0.05$ & 59, 60  \\
HD 217786 & \nodata & 13.0$\pm0.8$ & 1319 $\pm4$ & 0.40$\pm0.05$ & 1.02$\pm0.03$ &5966$\pm65$ & 
 4.35$\pm0.11$  & -0.135$\pm0.043$ & 61 \\
WASP-30  & $60.96\pm 0.89$ & \nodata &    4.156736$\pm0.000013$ &      0 (adopted) & $      1.17\pm    0.03$ &      6201$\pm       97$ &
      4.28$\pm     0.01$ & $   -0.03\pm      0.10$ & 13  \\
COROT-15  & $      63.3\pm       4.1$ & \nodata &  3.06036$\pm0.00003$ &      0 (adopted) & $      1.32\pm     0.12$ &      6350$\pm       200$ &
      4.3$\pm      0.2$ & $     0.10\pm      0.20$ & 14  \\
LHS 6343  & $      62.9\pm       2.3$ & \nodata &      12.71382$\pm0.00004$ &    0.056$\pm0.032$ & $     0.370\pm    0.009$ &     \nodata  &
      4.851$\pm     0.008$ & $    0.04\pm     0.08 $ & 15  \\
Corot-3b & $      21.66\pm       1.00$ & \nodata &      4.25680$\pm0.000005$ &      0 (adopted) & $      1.37\pm    0.09$ &      6740$\pm       140$ &
      4.22$\pm     0.07$ & $   -0.02\pm     0.06 $ & 19  \\
XO-3b  & $      11.8\pm      0.59$ & \nodata &      3.1915239$\pm0.0000068$ &     0.26$\pm0.017$ & $      1.213\pm    0.066$ &      6429$\pm       100$ &
      4.244$\pm     0.041$ & $    -0.177\pm     0.080$ & 20  \\
KOI-423b & 18.0$^{+0.93}_{-0.91}$ & \nodata  &      21.0874$\pm 0.0002$ &     0.121$^{+0.022}_{-0.023}$ &  1.1$^{+0.07}_{-0.06}$ &  6260$\pm140$ &
      4.1$\pm      0.2$ & $    -0.29\pm      0.10$ & 23   \\
KELT-1b &27.23$^{+0.5}_{-0.48}$   &  \nodata &      1.217514$\pm 0.000015$ &     0.0099$^{+0.01}_{-0.0069}$& $1.324\pm0.026$ &  6518$\pm 50$ &
      4.229$^{+0.012}_{-0.019}$ & $ 0.008\pm      0.073$ & 56  \\
KOI-205.01 & 35 & \nodata  &      11.72 &     \nodata &  \nodata &     5060 &
       4.57 &  -0.17 & 57  \\
KOI-554.01 & 80 & \nodata  &      3.66 &     \nodata &  \nodata &     5835 &
       4.64 & -0.08 & 57  \\
\enddata
\tablerefs{1 \citet{sahlmann11a}; 2 \citet{sato10}; 3 \citet{hatzes05}; 4 \citet{nidever02}; 5 \citet{benedict10}; 6 \citet{wittenmyer09}; 7 \citet{liu08}; 8 \citet{omiya09}; 9 \citet{moutou09}; 10 \citet{endl04}; 
 11 \citet{butler06}; 12 \citet{udry02}; 13 \citet{anderson11}; 14 \citet{bouchy11a}; 15 \citet{johnson11}; 16 \citet{lee11}; 17 \citet{halbwachs00}; 18 \citet{diaz12}; 19 \citet{deleuil08}; 20 \citet{winn08};  
 21 \citet{ma13}; 22 \citet{marcy01}; 23 \citet{Bouchy11b}; 24 \citet{galland06}; 25 \citet{gerbaldi07}; 26 \citet{perrier03}; 27 \citet{jenkins09}; 28 \citet{correia05}; 29 \citet{feroz11}; 30 \citet{zucker04}; 
  31 \citet{winn10}; 32 \citet{nie09}; 33 \citet{frink02}; 34 \citet{kane11}; 35 \citet{onehag09}; 36 \citet{lovis07}; 37 \citet{santos09}; 38 \citet{sato09}; 39 \citet{mug06}; 40 \citet{vogt00};  
   41 \citet{wittenmyer09}; 42 \citet{zucker00}; 43 \citet{latham89}; 44 \citet{kane09}; 45 \citet{fleming10}; 46 \citet{laf11}; 47 \citet{konopacky10}; 48 \citet{sah11}; 49 \citet{crepp11}; 50 \citet{wang12};  
   51 \citet{patel07}; 52 \citet{curto10}; 53 \citet{santos10}; 54 \citet{sozzetti06}; 55 \citet{siverd12}; 56 \citet{sozzetti10}; 57 \citet{santerne12}; 58 \citet{fleming12}; 59 \citet{fischer02}; 60 \citet{mar10}; 61 \citet{moutou11}; 62 \citet{ma14}; 63 \citet{jiang13}}
\end{deluxetable}

\clearpage

\end{document}